\documentclass[letterpaper]{article}
\usepackage{authblk}

\usepackage{geometry}
\usepackage{cite}
\usepackage{amsmath,amssymb,amsfonts}
\usepackage{algorithmic}
\usepackage{comment}
\usepackage{graphicx}
\usepackage{subcaption}
\usepackage{float}
\usepackage{todonotes}
\usepackage{textcomp}
\usepackage{xcolor}
\usepackage{multirow}
\usepackage{url}
\usepackage{amssymb}
\usepackage{tabularx}
\usepackage{soul}
\setstcolor{green}
\DeclareMathOperator*{\argmin}{arg\,min}
\def\BibTeX{{\rm B\kern-.05em{\sc i\kern-.025em b}\kern-.08em
    T\kern-.1667em\lower.7ex\hbox{E}\kern-.125emX}}

\newcommand{\etal}{\textit{et al}.}

\begin{document}

\providecommand{\keywords}[1]
{
  \small	
  \textbf{\textit{Keywords---}} #1
}

\title{A Review of Confidentiality Threats Against Embedded Neural Network Models\\}

\author[1,2]{Raphaël Joud}
\author[1,2]{Pierre-Alain Moëllic}
\author[1,2]{Rémi Bernhard}
\author[3]{Jean-Baptiste Rigaud}

\affil[1]{CEA Tech, Centre CMP, Equipe Commune CEA Tech - Mines Saint-Etienne, F-13541 Gardanne, France}
\affil[2]{Univ. Grenoble Alpes, CEA, Leti, F-38000 Grenoble, France\protect\\
\textit{\{raphael.joud,pierre-alain.moellic,remi.bernhard\}@cea.fr}}
\affil[3]{Mines Saint-Etienne, CEA Tech, Centre CMP, F - 13541 Gardanne, France\protect\\
\textit{jean-baptiste.rigaud@emse.fr}}

\date{}

\maketitle

\begin{abstract}
Utilization of Machine Learning (ML) algorithms, especially Deep Neural Network (DNN) models, becomes a widely accepted standard in many domains more particularly IoT-based systems. DNN models reach impressive performances in several sensitive fields such as medical diagnosis, smart transport or security threat detection, and represent a valuable piece of Intellectual Property. Over the last few years, a major trend is the large-scale deployment of models in a wide variety of devices. However, this migration to embedded systems is slowed down because of the broad spectrum of attacks threatening the integrity, confidentiality and availability of  embedded models. In this review, we cover the landscape of attacks targeting the confidentiality of embedded DNN models that may have a major impact on critical IoT systems, with a particular focus on model extraction and data leakage. We highlight the fact that Side-Channel Analysis (SCA) is a relatively unexplored bias by which model’s confidentiality can be compromised. Input data, architecture or parameters of a model can be extracted from power or electromagnetic observations, testifying a real need from a security point of view.
\end{abstract}


\section{Introduction}
\label{introduction}
\begin{figure*}[t]
	\centering
	\includegraphics[width=\textwidth]{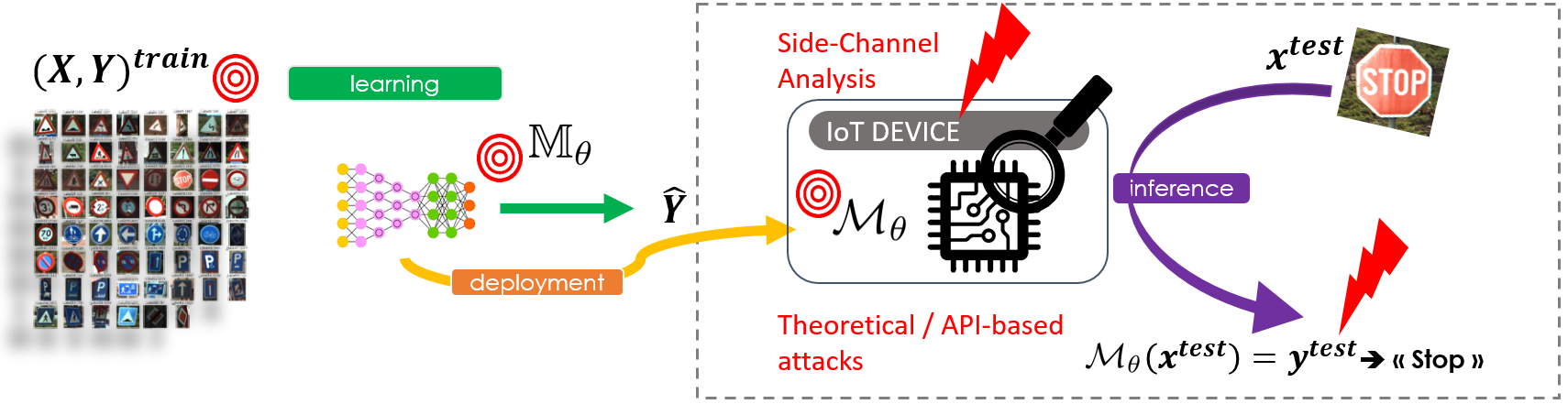}
	\caption{The traditional (supervised) ML pipeline and the scope of this review (gray dotted area). The threats concern the exploitation of algorithmic flaws as well as SCA techniques to extract model's information or exploit data leakage (red targets).} \label{attack_surface_scope}
	\vspace{-4mm}
\end{figure*}

An increasing number of edge applications can take advantage of the high effectiveness of Deep Neural Networks (DNN)
such as image classification or natural language processing. Based on that, deploying DNN models on hardware platforms catch more and more interest. Running DNN on edge devices offers many advantages. A first one is \textit{power consumption efficiency}, since high-frequency communication with a cloud-based ML services can be prohibitive. Model compression (e.g., quantization, pruning) and inference optimization (e.g., low bitwidth arithmetic) enable to keep reducing models processing consumption costs. 
Then, model's embedding solves \textit{latency} issues that may occur for applications that rely on important amount of data or need to work at high frequency. Another advantage is linked to \textit{privacy} and \textit{confidentiality} concerns by keeping locally critical information and avoid their transmission to potentially flawed networks.

These advantages explain the massive effort for the deployment of models on a wide variety of devices, especially for IoT applications. For now, this deployment is essentially focused on inference purposes since training still requires important computing capabilities. However, the concerned hardware platforms gather a broad range of integrated circuits, from constrained 32-bit microcontrollers, low-power processors, system-on-chip with HW accelerators to high performance FPGA.

\subsection{Attacks on the Machine Learning pipeline}
Among the major obstacles of the deployment of ML models, security issues are notably critical and concern the whole ML pipeline. The traditional ML pipeline gathers the \textit{learning step}, where training data are processed to optimize a (usually) parametric model such as a neural network and the \textit{inference step} where the learned model is used to achieve its task on unseen test inputs. An important amount of works raises major security flaws everywhere in the pipeline with threats targeting integrity (e.g., adversarial examples), confidentiality (e.g., model extraction) and availability (e.g., over-power consumption).%~\cite{sponge_examples}).

However, contrary to secure applications such as cryptographic modules, designed to be robust against theoretical attacks (i.e. cryptanalysis), embedded ML models suffer from a twofold attack surface: 
\begin{itemize}
\item \textit{Algorithmic attack surface}: several theoretical flaws have been successfully exposed such as the well-known adversarial examples that enable to fool a model with imperceptible perturbations of the inputs. This kind of attacks are regularly referenced as \textit{Application Programming Interface (API) -based} attacks and we use this notation in this work.
\item \textit{Physical attack surface}: being embedded in a device (hardware platform), a model can also be attacked by taking advantage of flaws related to its particular physical implementation, such as side-channel (SCA) and fault injection analysis (FIA).
\end{itemize}

This extended attack surface provides many possibilities to an attacker since it opens door towards two highly critical problems. First, combining the theoretical evidences from API-based attacks and side-channel analysis may reduce the complexity of a \textit{Model Extraction} task that consists in Intellectual Property theft, as Neural Network can be long, expensive to train and represent a real asset for its owner. On another hand, a successful model extraction attack changes the level of control an attacker has over the system. Attacker can then shift from a black-box context to an almost (if not complete) white-box one. This kind of threat can be dramatic as it would drastically ease the research for adversarial examples \cite{carlini2017towards}, \cite{su2019one}, membership inference attacks \cite{shokri2017membership} or data leakage exploitation \cite{carlini2019secret}. This context change is a critical risk regarding model’s integrity or user’s data confidentiality because the attacks previously mentioned keep high effectiveness when transferred from a model to a similar one.

\subsection{Side-channel analysis for embedded DNN models}

\textit{Physical attack surface} leverage model's implementation weaknesses, which can be exploited through Side-Channel Analysis (SCA). Every running program has a specific impact on physical quantities such as power consumption or electromagnetic emanations. SCA aims at exploiting these physical signatures to extract sensitive information related to program's execution.
Initially oriented against cryptographic devices, SCA appeared to be extremely powerful over the years and could be used against embedded DNN for model extraction.

Several techniques have been proposed to exploit side-channel leakages: 
\begin{itemize}
    \item \textit{Simple Power Analysis} (SPA) is only based on the visual interpretation of \textit{patterns} using a single or few observations (hereafter, called \textit{traces}). This analysis, also mentioned as \textit{Timing Analysis}, relies on some knowledge of the studied program.
    \item \textit{Differential Power Analysis} (DPA) is focused on the statistical dependencies between several physical measurements and some assumptions about the targeted program. For example, in key-recovery challenge, the assumptions are made by the attacker based on key hypothesis (e.g., 256 for one key byte) and a \textit{leakage model} such as the popular Hamming Weight or Hamming Distance models that link power consumption with the energy used to process or shift one bit. 
    \item \textit{Correlation Power Analysis} (CPA): it is a specific DPA implementation in which the statistical technique used to recover dependencies is the computation of Pearson Correlation Coefficient.
\end{itemize}
For more details about SCA, we refer interested readers to \cite{mangard_power_2007}.

\subsection{Contributions and scope}
This review aims at alerting to the criticality of confidentiality-based threats against embedded models by highlighting the case of black-box model reverse engineering that may jointly exploits API-based and physical attacks. For physical threats, we limit the scope of our review on SCA since it gathers the most significant efforts. However, Breier \etal \cite{sniff} pave the way to model extraction techniques exploiting fault attacks (bit flip model in \cite{sniff}). We illustrate the scope of the paper in Figure~\ref{attack_surface_scope} with a representation of the ML pipeline.
This document is organized as follow: after introducing some notations and formalism used throughout this work in Section~\ref{notation}, we detail the considered threat model in Section~\ref{threat_model}. Section~\ref{data_leakage} is dedicated to confidentiality issues with the exploitation of training data leakages. The last section (\ref{model_extraction}) proposes state-of-the-art perspectives regarding model extraction techniques.

\section{Notations and Formalism}
\label{notation}
We distinguish a DNN model, $\mathbb{M}$ as an abstract algorithm from its physical implementations $\mathcal{M}$. One model $\mathbb{M}$ (e.g., a convolutional neural network trained on ImageNet in order to classify colored images) may be deployed for inference purpose in a 32-bit microcontroller ($\mathcal{M}^1$), a computer with an advanced CPU  ($\mathcal{M}^2$) or a FPGA ($\mathcal{M}^3$). From a purely functional point of view, the models $\mathcal{M}^{*}$ rely on the same abstraction $\mathbb{M}$ but strongly differ in terms of implementation and their respective hardware environments. Moreover, because of some hardware constraints, the parameters of $\mathcal{M}$ may differ from $\mathbb{M}$ (e.g., weights quantization) and therefore the functionality equivalence is not possible. 

A supervised neural network model $\mathbb{M}_\theta$ is a parametric model that aims at mapping the input space $\mathcal{X}=\mathbb{R}^d$ (e.g., images, speech) to the output space $\mathcal{Y}$. %$\Theta$ is the set of parameters to be learned and $\mathcal{D}$ is the data distribution. 
For classification, $\mathcal{Y}$ is a finite set of labels. $\mathbb{M}_{\theta}$ is trained by minimizing a loss function $\mathcal{L}$ that quantifies the error between a prediction $\hat{y} = \mathbb{M}_\theta(x)$ and the ground-truth label $y$.  
\begin{comment}
\begin{equation}
\Theta^* = \argmin_\theta\big(
\mathop{\mathbb{E}}_{x,y\sim \mathcal{D}}\big[\mathcal{L}(\mathbb{M}_{\theta}(x);y)\big] \big)

\label{ml_definition}
\end{equation}
\end{comment}
At inference time, when feed by a input $x \in \mathcal{X}$, the learned model outputs a set of raw predictions (the \textit{logits}) that are mapped in a probabilistic formalism thanks to a softmax function.

\section{Threat Model}
\label{threat_model}
In this section, we detail the threat model used in this paper as a compulsory step to define properly the adversary's goals and capacities as well as the defender requirements.

\noindent\textbf{Attack surface.} As previously mentioned, the attack surface in this work is related to an adversary that has physical access to the target device in which a DNN model is implemented. 

\noindent\textbf{Adversary's goal.} The main objective of the attacker is to strike the confidentiality/privacy of the system. Therefore, he  may has several goals: 
\begin{itemize}
\item target the model by itself: extract the architecture of the model as well as all the parameters (see Section~\ref{model_extraction})
\item extract information related to the data (see Section~\ref{data_leakage}), predominantly the training data. 
\end{itemize}
We summarize these goals and targets in Table~\ref{target_goal_domain_table} with a set of representative critical IoT domains.
\begin{table}[t]
 \centering
 %\begin{tabular}{|c|c c|}
 \resizebox{\textwidth}{!}{%
  \begin{tabular}{ccc}
    \hline
    Target & Goal & Critical IoT domains \\
    \hline
    \hline
    \multirow{2}{*}{Model ($\mathcal{M} \rightarrow \mathbb{M}$)} & IP theft, steal performance, & Transport, Industry,\\
    & shift to white-box context & (Cyber)Security\\
    %& white-box context & (Cyber)Security\\
    \hline
     %& \multirow{4}{*}{Corrupt privacy} & Medical,\\
    %Training data, & \multirow{2}{*}{Corrupt privacy} & Medical, Biometry,\\
    %Membership & & Defense, Finance\\
    %\hline
    Training data, Membership & Corrupt privacy & Medical, Biometry, Defense, Finance\\

    \hline
  \end{tabular}}
  \caption{Summary of the targets and goals of a confidentiality attack against an embedded DNN with representative related IoT domains.}
 \label{target_goal_domain_table}
 
\end{table}
\noindent\textbf{Adversary's capacity and knowledge.} The more realistic setting is to consider a black-box model, i.e. the adversary has access neither to the internal parameters of the models nor to its architecture.

\begin{figure}[t]
	\centering
	\includegraphics[scale=0.15]{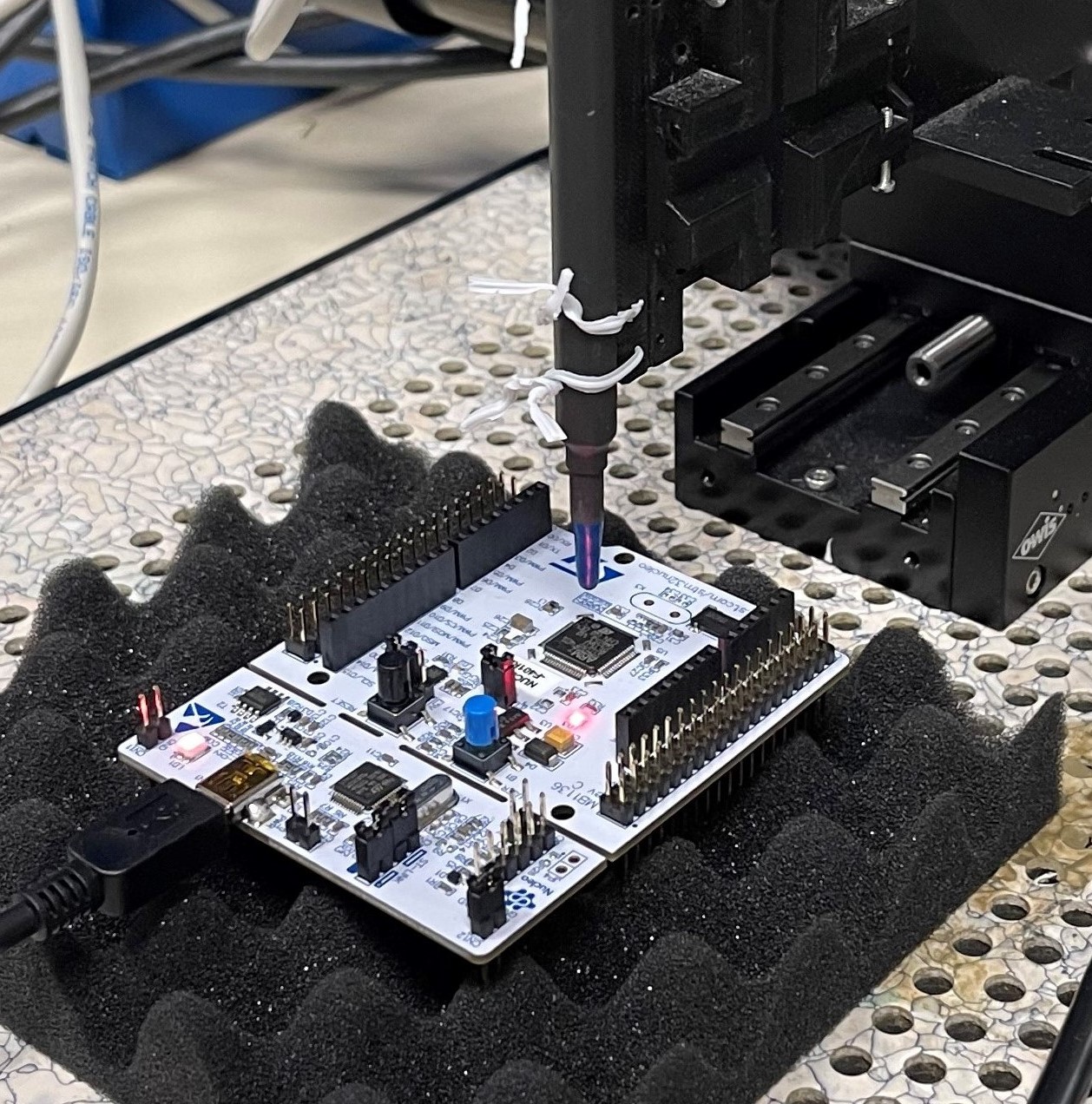}
	\caption{Classic setup for electromagnetic emanations recording (target is a Cortex M4 32-bit microcontroller.} 
	\label{illustration_sca_setup}
\end{figure}

On the contrary, the adversary can query the device by feeding the model with input data and collect the corresponding outputs. Note that these outputs may be minimal information (only a label) or complete one (e.g., all the prediction probabilities). Moreover, the adversary can perform side-channel analysis on the device (or by using a clone of the device) by observing the power consumption or the electromagnetic emanation of the target thanks to a probe (see a typical SCA set-up illustration in Figure~\ref{illustration_sca_setup}).  
We assume that the attacker targets the inference. Then, he/she has no impact on the training process neither on the deployment phase. However, according to the nature and the context of the task performed by the model, the adversary may have some assumptions on the training data and even on the nature of the model (e.g., computer vision tasks are likely to use convolutional neural networks or model language recurrent neural networks).  
In the following sections, according to this threat model, we discuss the different adversary's goals by highlighting state-of-the-art results on both theoretical flaws that \textit{directly} exploit input/output pairs and assumptions about the (black-box) target model $\mathbb{M}$; and physical attacks (SCA) that exploit physical observation of the implementation $\mathcal{M}$.

\section{Data Leakage}
\label{data_leakage}
\subsection{Extraction of Training Data}

DNN tend to encapsulate information about the training data that may leak \textit{intrinsically} from the model (i.e. through the exploitation of the model's parameters) or through the model's outputs (labels or  scores). Carlini \etal~\cite{carlini2019secret} highlight critical \textit{unintended memorization} issues for language models trained on a large-scale text data set. Such issue may be solved by Differential Privacy (DP)~\cite{dwork2014dp}, which aims at limiting the information one may derive from the output about the input. 

Another critical threat is known as \textit{attribute inference attack} (or \textit{model inversion})~\cite{yeom2018privacy} and mainly concerns applications where input data includes \textit{non-sensitive} features (e.g., public knowledge) and some \textit{sensitive} features that are not available for the adversary. The goal of the attack is to extract the value of these sensitive features with the knowledge of the model output and the \textit{non-sensitive} features.
Interestingly, a SCA experiment has been proposed in~\cite{wei_i_2018} for a pure model inversion use case, by recovering MNIST (grayscale image of digits) images from the inference power traces of a CNN deployed on a FPGA-based accelerator.

\subsection{Membership Inference}

The goal of an adversary mounting a Membership Inference Attack (noted MIA hereafter) against a model $\mathbb{M}$ is to decide whether or not an example $x$ was used or not at training time.
For example, considering a model trained with health information gathered from people suffering from a particular disease, this type of attack therefore results in the disclosure of the identity of some people having this disease.

Yeom \etal~\cite{yeom2017privacy} introduce two simple MIAs, which require the adversary to know at least the loss value. The first one is the the "0-1" baseline attack: an example is decided to belong to the training set if it is well predicted.
The second one ("Loss-attack") directly links the probability to belong to the training set to the loss value.

Advanced MIA principles have been proposed by \cite{shokri2017membership} with an adversary having access to the confidence output scores and a set of data closed to the ones used for the training. The idea is to train a \textit{shadow} model $\mathbb{M}^s_c$ for each possible label class $c$. The adversary uses each $M^s_c$ to build a set $((c,\hat{y}), in / out)$, by querying train ($in$) or test ($out$) points used to train $M^s_c$, to get the confidence vector $\hat{y}$. This set, consisting of confidence score vectors labeled by their membership to the train or test is then used to train an \textit{attack model}, taking as inputs the $\hat{y}$ and as outputs the $in$ or $out$. \cite{salem2018ml} shows that this attack is also feasible when relaxing assumptions on the strength of the adversary, keeping high efficiency rate. 

To mitigate this type of attacks, many defense schemes aim at adding noise to the confidence output vector \cite{nasr2018machine, Jia2019, xue2020use}.
However, these protections have been thwarted by so-called "Label-Only" MIAs, which rely only on the output label \cite{choo2020labelonly}. %, li2020labelleaks}.
The efficiency of this attack compared to white-box attacks has already been observed experimentally \cite{song2020systematic}. Moreover, the optimal strategy to perform MIAs has been formally derived, and it shows that an adversary only needs to have access to the classification loss value \cite{sablayrolles2019}. Therefore, the white-box setting does not bring any advantage in terms of MIA success.

Many efforts have been devoted to identify factors that influence the success of MIAs \cite{tonni2020data}: the size of the training set, the number of classes and the intra-class standard deviation, the mutual information between outputs and inputs, or the amount of generalization.
However, all these factors are just a cause for the reduction of the generalization capacity~\cite{song2019privacy}. Thus, \cite{li2020membership} proposes to reduce the generalization gap (difference between train and test set loss) by adding to the training loss a measure of the performance gap between training and validation samples. This simple procedure aims at minimizing the difference between probability scores for training and validation examples.

Another line of defense against MIAs focuses on Differential Privacy. However, such defenses face a classical issue with DP since it has been observed that for a certain privacy level needed to thwart MIAs, the accuracy of the target model was decreased in a non-acceptable way \cite{rahman2018membership}.

\section{Model Extraction}
\label{model_extraction}
\begin{figure}[t]
	\centering
	\includegraphics[width=0.75\linewidth]{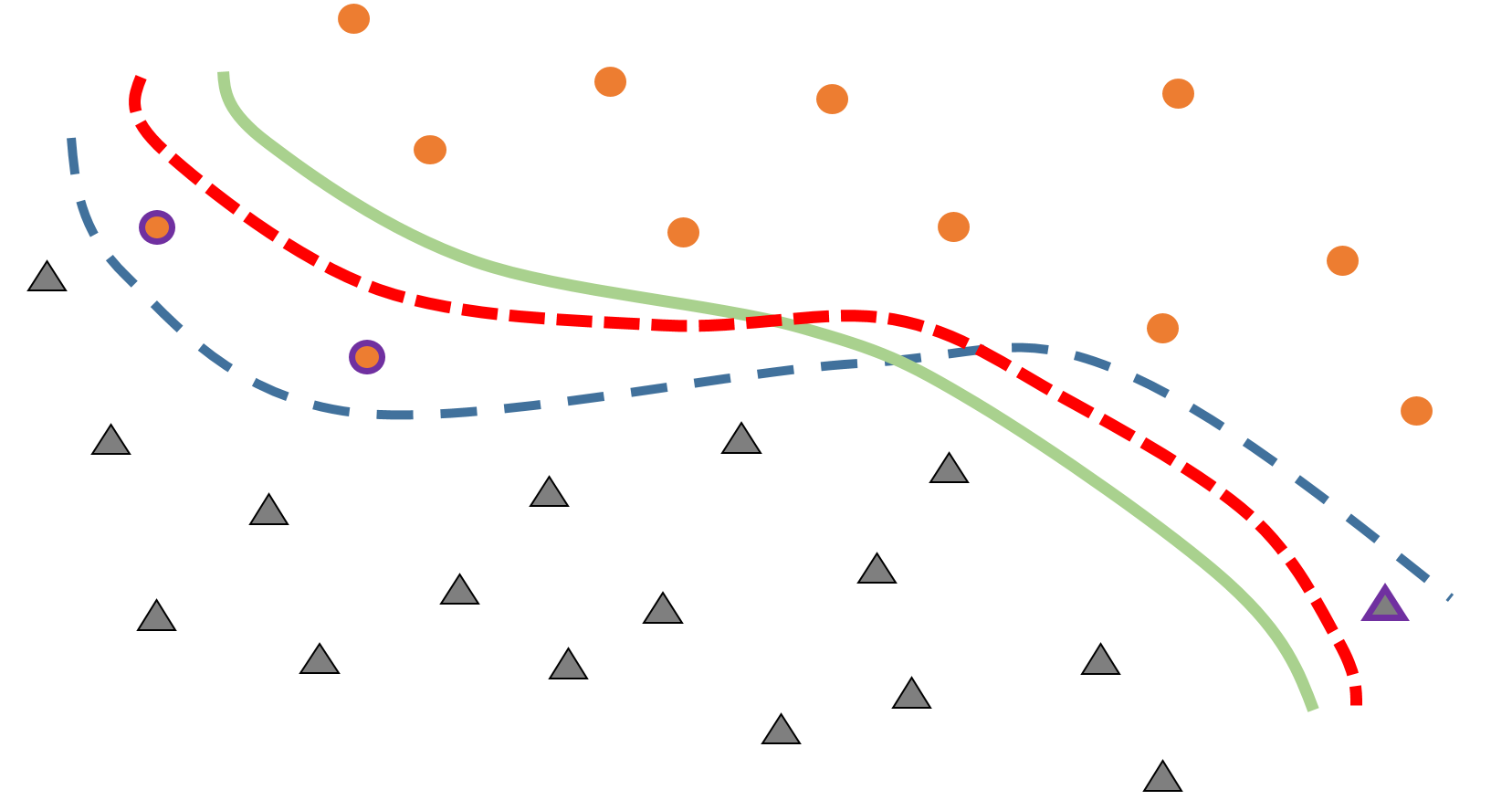}
	\caption{Illustration from~\cite{jagielski_high_2020}. Original model's decision frontier is modeled with the green line. Red-dashed line is the \textit{Fidelity}-oriented reconstruction as it matches predictions in the same manner as the original green curve. Whereas blue-dashed line is an \textit{Accuracy}-oriented approach: incorrect prediction from green curve are corrected.} \label{fidelity_accuracy}
\end{figure}

Being an important intellectual property, a DNN model is a privileged target that an adversary aims at reverse engineering especially if the black-box model is insufficiently protected. In the following part, we present different shades of model extraction attacks starting with algorithmic-oriented techniques. Then, we broach SCA-based attacks. All the attacks mentioned in the following part are gathered in table \ref{sum_table}.

\begin{table*}[t]
 \centering
 %\begin{tabular}{|c|c c c c|}
 \resizebox{\textwidth}{!}{%
 \begin{tabular}{ccccc}
  %\hline
  %\multicolumn{5}{|c|}{Model Extraction Attack} \\
  \hline
  Attack & Implementation & $\mathbb{M}$ & $\mathcal{M}$ & Specificity\\
  \hline
  \hline
  Carlini \textit{et al.} \cite{carlini_cryptanalytic_2020} & N.S. & \checkmark & & For ReLu Multi Layer Perceptron\\
  Jagielski \textit{et al.} \cite{jagielski_high_2020} & N.S. & \checkmark & & For ReLu Multi Layer Perceptron\\
  Oh \textit{et al.} \cite{oh_towards_2018} & N.S. & \checkmark & & ML model able to recognize other models\\
  \hline
  Batina \textit{et al.} \cite{batina_csi_2019} & µC & & \checkmark & SCA\\
  Regazzoni \textit{et al.} \cite{regazzoni_machine_2020} & µC & & \checkmark & SCA\\
  Maji \textit{et al.} \cite{maji2021leaky} & µC & & \checkmark & SCA \\
  Xiang \textit{et al.} \cite{xiang_open_2020} & µC & \checkmark & \checkmark & SVM to classify DNN architecture from SC traces \\

  \hline
  
  Hua \textit{et al.} \cite{hua_reverse_2018} & FPGA & & \checkmark & Memory-access pattern and timing SC\\
  Dubey \textit{et al.} \cite{dubey_maskednet_2020} & FPGA & & \checkmark & CPA and advanced leak-model\\
  \multirow{2}{*}{Yu \textit{et al.} \cite{yu_deepem_2020}} & \multirow{2}{*}{FPGA} & \multirow{2}{*}{\checkmark} & \multirow{2}{*}{\checkmark} & EM SC traces (model's architecture) \\
  & & & & and AE to train substitute model\\
  \hline
  Breier \textit{et al.} \cite{sniff} & N.S. & & \checkmark & Fault Injection Analysis (FIA)\\
  \hline
 \end{tabular}}
 \caption{Summary table of model extraction attacks mentioned in this work and their specificity. Abbreviations: NS = Not Specified, µC = Microcontroller, $\mathbb{M}$ refers to API-based attacks and $\mathcal{M}$ to Physical Attacks (e.g., SCA)}
 \label{sum_table}
\end{table*}

Most of published works using side-channel analysis against embedded neural networks are focused on model extraction. However, it is important to distinguish what kind of extraction is to be performed. Jagielski \etal~\cite{jagielski_high_2020}~define a clear difference between attacks targeting \textbf{Fidelity} and \textbf{Accuracy}:
\begin{itemize}
    \item Fidelity measures how well extracted model predictions match those from the original model (victim target of the extraction attack).
    \item Accuracy aims at performing well over the underlying learning task of the original model based on a testing data-set.
\end{itemize}

Figure~\ref{fidelity_accuracy} illustrates these concepts (from~\cite{jagielski_high_2020}).

\subsection{API-based model extraction attacks}

According to previous notions, it is possible to distinguish different goals regarding extraction strictly speaking. An adversary may want to precisely extract model's characteristics in order to obtain a kind of clone. By achieving this goal, the attacker tremendously enhances his level of control over the system allowing to go from a black-box context to a white-box one. Attacker would then be able to easily design other attacks harming model's integrity and data confidentiality.
On the other hand, attacker might want to steal the ``performance" of an already-trained model for least possible effort without paying much interest to the exact value of model's characteristics. Indeed, high-performance neural network can take a significant amount of time and investment to be trained and deployed. Simply stealing its performance would avoid the attacker to go through a long and expensive training phase and harm the individuals or company behind the original model. This kind of attack is largely underrepresented in model extraction related publications. However, in~\cite{jagielski_high_2020}, Jagielski~\etal take an interest about these attacks with the idea that model predictions are more informative than training data labels. They compare the performance achieved when using a fully-supervised learning or semi-supervised learning upon part of the original training data set. Performance reached when using semi-supervised learning are significant and according to the publication, could be further improved by combining active-learning techniques with semi-supervised learning.

Considering fidelity-oriented model extraction attacks in the black-box context, the attacker can interact with a reduced number of elements. Generally, he can provide inputs to the model and receive model's prediction, then the adversary could be able to craft special inputs that would reveal highlighted information. Based on this idea, \cite{jagielski_high_2020} work on models using ReLU activation function in order to use specially crafted inputs that put exactly one neuron of the hidden layer in a critical state. This critical state forces one neuron to change sign and allows to identify the weight associated to it. By doing this for every neurons, the authors are able to reconstruct the entire weight matrix of the hidden layer, noting that models involved in this work are structured with only one hidden layer. This methodology is extended to models with more than one hidden layer in \cite{carlini_cryptanalytic_2020} by considering the model extraction as a cryptanalytic problem with a differential attack.

Chosen or crafted-query approach is also exploited in \cite{oh_towards_2018} but the attack relies on a \textit{metamodel} trained to predict the attributes of a black-box model, as a \textit{classifier of classifier}. Working with MNIST classifiers, different versions of this \textit{metamodel} are developed and combined in a way that, in the end, the \textit{metamodel} is able to reveal information about model's architecture and hyperparameters even with restricted output information (label-only) or difficult black-box contexts.
 
\subsection{Side-Channel Analysis for Model Extraction}

From a pre-trained DNN model, an attacker with a fidelity-based objective would want to extract the model's architecture, the parameters or the activation functions.

Fidelity extraction attack can benefit from the information provided by hardware implementation, especially revealed through SCA. Moreover, context restrictions associated to purely software and mathematical extraction can be diminished by using physical attacks. In \cite{batina_csi_2019}, no specification is imposed to the model topology or activation function. They propose an attack based on electromagnetic emanations emitted by a microcontroller that allows to re-construct a model. The attack is composed of three main steps: \textit{1)} extraction of weights associated to each neurons with method similar to CPA with hypothesis about value of the weight; \textit{2)} distinction of the layer to which the neuron belongs with SCA techniques; \textit{3)} guessing of the activation function based on timing analysis. This procedure is repeated layer after layer to recover the entire architecture. This attack is also extended to CNN in the paper, requiring only small changes. A similar methodology is evoked in \cite{regazzoni_machine_2020} and presents comparable results.

In a recent work, Maji~\etal~\cite{maji2021leaky} demonstrate that even simple power and timing analysis enable to recover an embedded neural network model with different precision (floating point, fixed point and binary models) on three microcontroller platforms (8-bit ATmega328P, 32-bit ARM Cortex-MO+ and 32-bit RISC-V RV32IM). Interestingly, to thwart the timing side-channel leakages, they propose two countermeasures based on a constant-time implementation of the ReLU operation (for 8-bit input) and a ad hoc floating-point representation.   

From the observation that embedded neural networks can also be based on already-existing and popular architectures (such as MobilNet, AlexNet or ResNet), Xiang~\etal~\cite{xiang_open_2020} propose to identify the structure of models deployed on a Raspberry Pi using power consumption traces. The identification, performed by a Support Vector Machine classifier, is allowed by the fact that each architecture is a combination of different layers having their own consumption profile.

Most of the works related to model extraction are focused on CNN embedded on FPGA accelerator. In \cite{hua_reverse_2018}, the authors use the fact that CNN states and parameters are stored in off-chip memory due to their size. They use memory-accesses and timing side-channel to reveal information like architecture (number of layers, data dependencies among layers, size of features maps and kernel filters) or weights from AlexNet and SqueezeNet. %Observing memory-access patterns during inference allows to obtain information like number of layers, data dependencies among layers, and size of features maps and kernel filters. 
Weights extraction is based on utilization of functions that dynamically prune zeros resulting from ReLU activation function. With use of such function, a change in the number of non-zero values in the Output Feature Map (OFM, output of a layer) reveals that a value from the Input Feature Map (IFM, input of the layer) has crossed the 0 boundary. By manipulating content of IFM and controlling apparition of non-zero values in OFM, attacker can recover weights characterizing a layer.

\cite{dubey_maskednet_2020} uses CPA to extract weights and bias from hardware implementation (Adder Tree) by focusing on switching activity of registers. The authors propose an advanced leakage model based on Hamming Distance. 
For example, the registers of the Adder Tree may contain the result of multiplication operation between a secret weight and a known input. 
An important part of this work presents the implementation of a countermeasure focused on masking all intermediate computations and occurs at different level of the model's implementation (Adder Tree, Activation Function ...).

In~\cite{yu_deepem_2020}, Yu~\etal~demonstrate a joint exploitation of SCA and a typical model extraction technique. They first extract the architecture of a black-box binary neural network model embedded on FPGA using electromagnetic traces. Then, a substitute model with the extracted structure is trained with a data set built from prediction returned by the victim model when specifically-crafted adversarial examples were passed to it.

\section{Conclusion}
\label{conclusion}
This paper proposes a review of the weaknesses that can critically threaten the confidentiality of neural network models deployed in typical IoT devices. Because the attack surface encompasses theoretical (algorithmic) flaws \textit{and} physical implementations threats, we particularly highlight two major issues~--~model extraction and training data leakage~--~that are representative to the use of side-channel analysis. We claim that model extraction is a major threat, with powerful attacks that have been recently demonstrated. Not only a model extraction enables to steal IP but also shifts an adversary into a harmful white-box paradigm. By knowing the internals of a model, an attacker may craft more advanced privacy-based attacks or integrity threats such as adversarial examples. For now, most \textit{fidelity}-oriented model extraction attacks are focused on a non-material attack perspective and few works are dedicated to typical IoT platforms (such as 32-bit microcontroller), mainly by using side-channel analysis to extract the architecture of a target model. More attention has to be paid about the joint exploitation of these techniques that may lead, in a very near future, to powerful attacks for a wide range of state-of-the-art models.

\section*{Acknowledgments}
This work is a collaborative research action that is partially supported by (\textbf{CEA-Leti}) the European project ECSEL InSecTT\footnote{\url{www.insectt.eu}, InSecTT: ECSEL Joint Undertaking (JU) under grant agreement No 876038. The JU receives support from the European Union’s Horizon 2020 research and innovation program and Austria, Sweden, Spain, Italy, France, Portugal, Ireland, Finland, Slovenia, Poland, Netherlands, Turkey. The document reflects only the author’s view and the Commission is not responsible for any use that may be made of the information it contains.} 
and by the French National Research Agency (ANR) in the framework of the \textit{Investissements d’avenir} program (ANR-10-AIRT-05, irtnanoelec);  and supported (\textbf{Mines Saint-Etienne}) by the French funded ANR program PICTURE (AAPG2020).

\bibliography{bib/biblio_1, bib/biblio_2, bib/biblio_3}
\bibliographystyle{ieeetr}

\end{document}